# Remote sensing of soil moisture using Rydberg atoms and satellite signals of opportunity

Darmindra Arumugam[1]*, Jun-Hee Park[1], Brook Feyissa[1]†, Jack Bush[1]†, Srinivas Prasad Mysore Nagaraja[1]†

[1]Jet Propulsion Laboratory, California Institute of Technology, Pasadena, 91109, California, USA

*Corresponding author. E-mail: darmindra.d.arumugam@jpl.nasa.gov;
Contributing authors: junhee.park@jpl.nasa.gov; brook.feyissa@jpl.nasa.gov; jack.d.bush@jpl.nasa.gov; srinivas.prasad.mysore.nagaraja@jpl.nasa.gov;
†These authors contributed equally to this work.

**Abstract**

Spaceborne radar remote sensing of the earth system is essential to study natural and man-made changes in the ecosystem, water and energy cycles, weather and air quality, sea level, and surface dynamics. A major challenge with current approaches is the lack of broad spectrum tunability due to narrow band microwave electronics, that limit systems to specific science variable retrievals. This results in a significant limitation in studying dynamic coupled earth system processes such as surface and subsurface hydrology, where broad spectrum radar remote sensing is needed to sense multiple variables simultaneously. Rydberg atomic sensors are highly sensitive broad-spectrum quantum detectors that can be dynamically tuned to cover micro-to-millimeter waves with no requirement for band-specific electronics. Rydberg atomic sensors can use existing transmitted signals such as navigation and communication satellites to enable remote sensing. We demonstrate remote sensing of soil moisture, an important earth system variable, via ground-based radar reflectometry with Rydberg atomic systems. To do this, we sensitize the atoms to XM satellite radio signals and use signal correlations to demonstrate use of these satellite signals for remote sensing of soil moisture. Our approach provides a step towards satellite-based broad-spectrum Rydberg atomic remote sensing.

Remote sensing of the earth system from space relies on a vast network of technologies[1]. Spaceborne radars play a key role for remote sensing applications spanning several science focus areas[2] such as surface, topography, vegetation science[3], or planetary boundary layer science[4]. Major challenges with state-of-art classical radar remote sensors today include band specific antennas and RF electronics, lack of tunability, and large form-factors that limit applicability to a specific science objective[2-4]. As a result, many distinct satellite radars covering different bands are needed to study coupled variables of the Earth system such as precipitation and soil moisture, which would require remote sensing covering long-to-short microwave wavelengths (I/P-K bands)[2]. To address this problem, a tunable radar system is needed that does not rely on band-specific microwave electronics.

Radar systems have traditionally required an onboard transmitter and receiver; However, increasingly remote sensing is achieved via use of existing satellite signals referred to as signals of opportunity (SoOp)[5-8], removing the need for an onboard transmitter. Dynamically tunable receivers could use radio reflectometry techniques[9-10] with SoOp signals spanning I-K bands to obtain wide-spectrum responses of the earth system. Downlink satellite radio signals from communication/navigation satellites could permit thousands of active transmissions spread throughout the radio window (VHF-to-K, example 137/260/360MHz /1.5/2.3/3.9/12.4/18.5/20.7GHz, see Supplementary Section 1) to be used for remote sensing. However, due to bandwidth limitations in antennas[11] and microwave electronics[12], traditional radio receivers are not practical to cover the entire radio window.

Atomic sensors use highly coherent quantum systems to probe atoms and measure weak signals with high sensitivity or precision[13]. Alkali atoms with high vapor pressure (such as Cesium/Cs and Rubidium/Rb) driven to a high principal quantum number and in the Rydberg states have been shown to be sensitive to microwave-to-millimeter waves[14-16]. Typically, two or more lasers are used to prepare and probe the atoms[15,16] to sense microwave fields, by first creating transparency of the probe laser light via electromagnetically induced transparency (EIT)[17], and then observing the perturbation of the EIT spectrum due to the external microwave fields. State-of-art techniques with Rydberg atomic systems can detect both amplitude and phase of the incoming microwave field[18] using a local oscillator (LO) to drive the atoms via super-heterodyning[19].

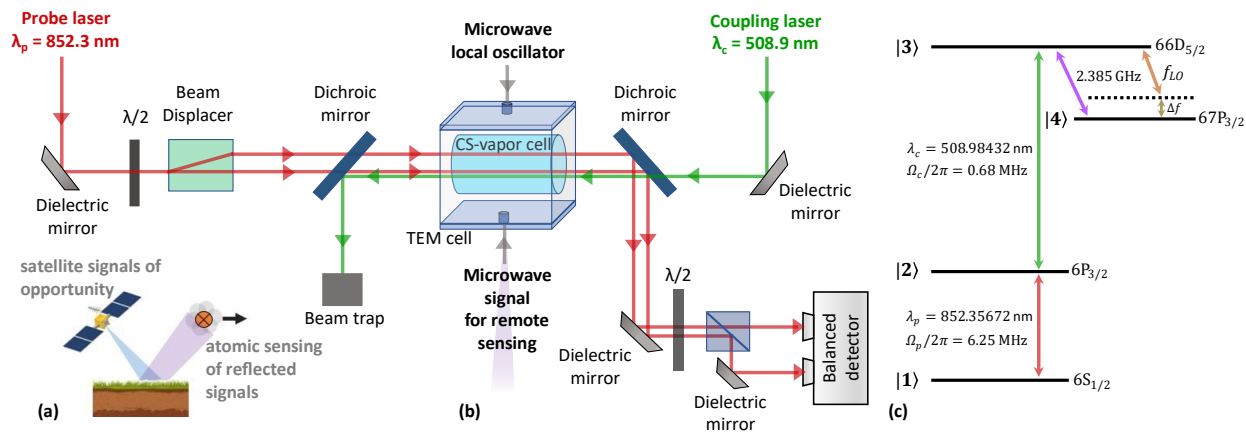

Fig. 1: (a) High level concept for remote sensing of soil moisture based on reflected satellite signals of opportunity sensed via Rydberg atoms. (b) The probe and coupler laser light are counter-propagated in a Cesium vapor cell to excite atoms to the Rydberg state. The vapor cell is located inside a transverse electromagnetic (TEM) cell, which is used for uniform microwave (MW) field coupling to the vapor cell. Both the MW signal to be detected and the local oscillator (LO) field is coupled into the TEM cell. A beam displacer is used to develop a reference probe beam for balanced detection to reduce technical noise. Inset shows a photo of the TEM/vapor cell. (c) The coupler drives the atoms to a principal quantum number of n = 66, and MW signal and LO is off-resonant by about 40-65 MHz from the next nearest state $66D_{5/2}$-$676P_{3/2}$ transition of 2.385 GHz.

We demonstrate use of Rydberg sensors configured with atomic super-heterodyning to remotely sense soil moisture using XM radio satellite signals (2.320-2.345 GHz). To achieve this, we utilize signal correlators with the use of a direct and reflected signal below the atomic readout noise floor and invert the magnitude of the correlator to soil moisture. Our approach opens a pathway to enable broad-spectrum remote sensing of the earth system using thousands of navigation and communications satellite transmissions.

## Results

### Correlations to extract modulated signals from noise

Soil moisture retrievals with existing satellite signals are most sensitive to microwave (MW) remote sensing in the P-S band[5] (Supplementary Section 1). We focus on the XM satellite radio band spanning 2.320-2.345 GHz[20]. Fig. 1a shows the high-level concept, where SoOp signals reflected of the ground are sensed by a Rydberg atomic sensor and interpreted for soil moisture. Fig. 1b shows the quantum optics setup developed for this purpose, which consists of Cesium atoms in a vapor cell at room temperature, laser systems comprised of a probe (~852nm) and tunable coupler (~509nm) laser, and microwave injection system. The probe and coupler are counter-propagated, and a reference probe beam is formed using a beam displacer to reduce technical noise. A transverse electromagnetic (TEM) cell is used to couple MW signals to the vapor cell with high efficiency. The probe, coupler Rabi frequency was set to $\Omega_{p,c}/2\pi$ = 6.25 MHz, 0.68 MHz, the laser wavelengths to drive $6S_{1/2}$

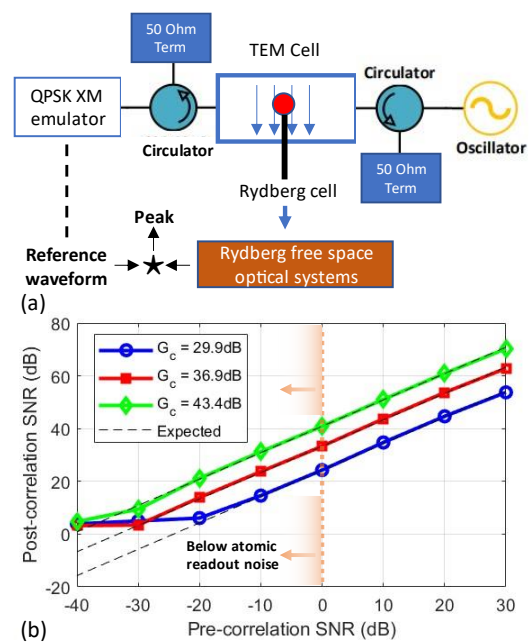

Fig. 2: A QPSK (quadrature phase shift key) signal with known bitstream is injected into the TEM/vapor cell to study signal-to-noise (SNR) after correlation with a reference waveform (a). The post correlation gains from the atomic readout demonstrated a gain close to the calculated $G_c$ (correlator gain) (by -3dB) (b).

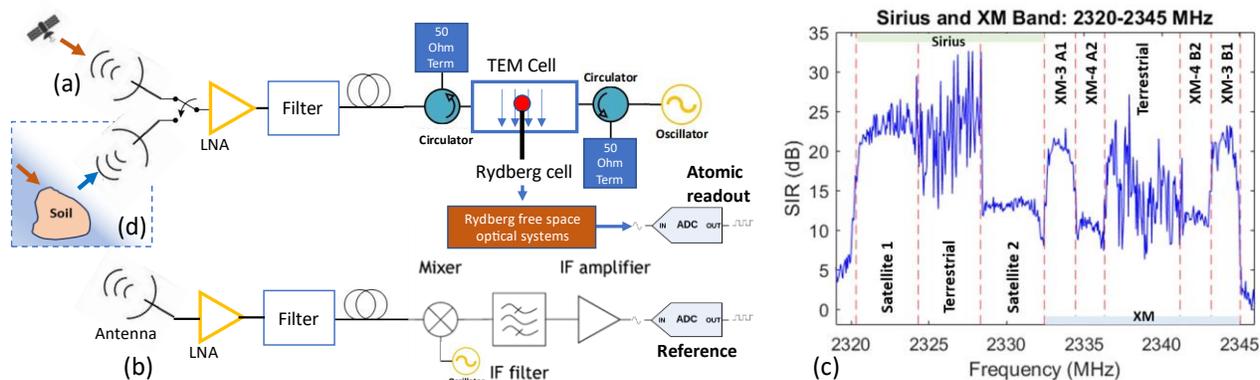

Fig. 3: A horn antenna is connected to a low noise amplifier (LNA), then a filter, and a long coaxial cable (50m long) to drive the atomic system via the TEM cell. (a) The horn is pointed towards the XM satellite (see Methods) to detect XM satellite signals. (b) A classical system with the same horn, LNA, filter, and coax cable is used to drive an RF mixer, IF (intermediate) filter and amplifier. Both quantum and classical readouts are digitized with an analog-to-digital convertor (ADC) and corelated to compute the envelop by sweeping local oscillator frequencies (c) Continuos readout of the XM satellite band using the correlations between (a,b). Both satellite and terrestrial repeaters are sensed. (d) The quantum systems (antenna) is directed downward towars the specular reflection to sense reflections off the soil sample (see Methods).

-6P$_{3/2}$-66D$_{5/2}$, and a strong (-19dBm) MW local oscillator (LO) (see Fig. 1)[19] is used to drive to 67P$_{3/2}$. The passband of the TEM cell was from DC-3 GHz (<3dB insertion loss) (details of experimental setup is presented in Methods).

To sense and retrieve soil moisture from heavily modulated and weak satellite signals, classical SoOp systems utilize signal correlators to extract signals out of the system noise floor[9] (see Supplementary Section 1). A similar approach for signal correlation is needed for the atomic system, because the SoOP signals are expected to be buried in the atomic readout noise (see Supplementary Section 1). We investigate this by injecting a QPSK (quadrature phase shift keying) modulated signal (10kHz modulation rate) into the TEM/vapor cell and correlate with a copy of the reference waveform (known random bit-stream) with $\Delta f$ = 50 MHz (see Fig. 1 and 2). The correlator gain (the processing gain from correlation) is $G_c = B\tau$, where $B$ is the modulated signal bandwidth, and $\tau$ the correlation time. Typical SoOp signal processing approaches are limited by the decorrelation time of the geophysical environment observed, resulting in $G_c$ between 30-43 dB (Supplementary Section 1). We estimate the pre-correlation signal-to-noise ratio (SNR) using a 1Hz resolution bandwidth power spectrum and compare to the post-correlation SNR obtained by peak detection[21]. We find that the correlator output SNR is higher than the pre-correlation SNR by $\sim 10 \log_{10} G_c$ (see Supplementary Section 2 for details), and that signals up to this value below the atomic readout noise floor can be extracted with high linearity (Fig. 2b). This demonstrates that weak modulated signals below the atomic readout noise can be extracted via correlation with a known reference waveform by as much as the correlation gain.

**Continuous XM satellite spectral envelop detection**
Recent work has demonstrated the first detection of XM satellite signals in a discontinuous spectrum via a Rb Rydberg atomic system[22]. Resonantly (or close to resonant) coupling with Rb systems would require n>90 and is generally impractical. As a result, the approach required use of a mm-wave (MM) source (119 GHz) to drive the Rb atoms to a high orbital angular momentum, $L > 4$, to sensitize the atom to lower MW frequencies[22]. To avoid high-L, which results in lower sensitivity to MW's and an additional MM source, we use Cs atoms driven to 66D$_{5/2}$ that gets close to a resonance ($\Delta f$ = 40-65 MHz, see Fig. 1c). The nearest state is 67P$_{3/2}$ with a resonance at 2.385 GHz. To continuously detect the entire XM satellite band (2.320-2.345 GHz), we configure a quantum-classical correlator system as shown in Fig. 3. A horn antenna (~10dB gain) is used to couple incoming MW to the TEM/vapor cell via a low noise amplifier (LNA, 31.5dB gain), bandpass filter, and 12m coaxial RF cable (~15dB loss) (Fig. 3a). A reference signal is obtained using the same components (antenna, LNA, filter, long coax) with a RF mixer (Level 13 with <6dB conversion loss), IF (intermediate) filter and amplifier (Fig. 3b). Both quantum and classical (reference) output is digitized by an analog-to-digital convertor (ADC) and correlated similar to Fig. 2. Both channels use a common LO, with attenuators used to set to -19dBm for the Rydberg LO and +13dBm for the classical reference LO. LO frequency was swept from 2.319-2.346 GHz to readout the entire spectrum via off-resonant Rydberg excitation[23]. Fig. 3c shows the output of the correlator after continuously sweeping the LO through the entire band. The result is the entire XM signal spectral envelop to include satellite and terrestrial repeaters. The Sirius satellite occupies the lower 12.5MHz band with 2 satellite sub-channels (~4MHz bandwidth/channel), while XM occupies the higher 12.5MHz band with 4 satellite sub-channels (~2MHz bandwidth/channel). We find the correlations to have a signal-to-interference (SIR) of >10dB and maximum of ~25dB for the satellite signals. The instantaneous bandwidth (throughout the band) was observed to be about 150kHz. See Methods for details.

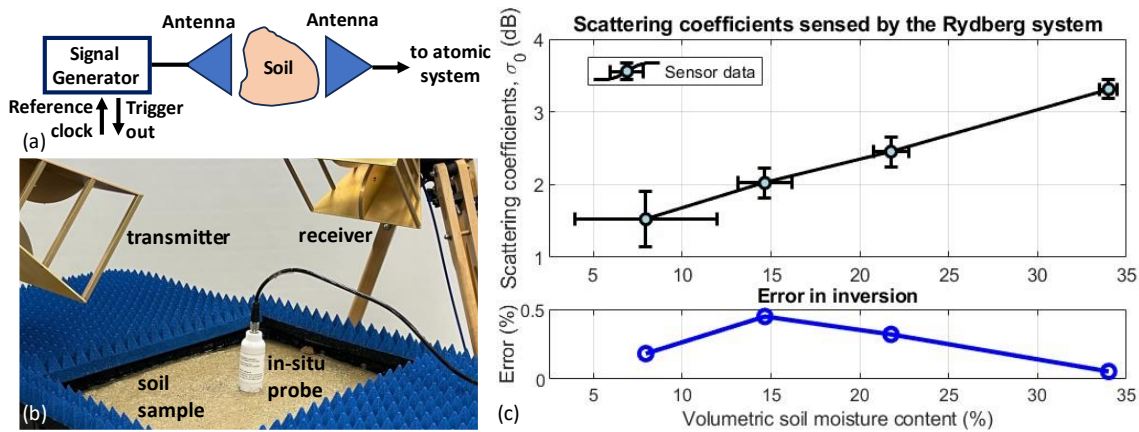

**Fig. 4: An emulated XM signal (2 MHz bandwidth QPSK signal with random bitstream) is used as a transmitter to study sensitivity of Rydberg atomic detectors to soil moisture (SM). (a) The transmitter is driven by a vector signal generator. (b) A soil sample (sand, fine particle size <0.42mm) is used in a sample container of 1.5x2x1.5 ft. MW absorber foam is placed outside the specular region to attenuate reflections of the container and ground. An in-situ SM probe is used to obtain ground truth measurements. (c) Radar scattering coefficients sensed by the Rydberg atomic readout (top) and error is SM (bottom) inversion as a function of volumetric SM content, showing <0.5% error in SM inversion (see Methods for details).**

**Soil moisture retrievals using emulated XM signals**

To study the sensitivity to soil moisture (SM), the system in Fig. 3d was used first in a laboratory environment with an emulated XM QPSK modulated signal (center frequency of 2.333GHz, 2MHz bandwidth, with known random bitstream, output power of -10dBm, no LNA used, and 12m cable loss of about 15dB). A vector signal generator (VSG) was used to generate the emulated signals, which was driven to a transmit antenna (see Fig. 4a, b), while a receive antenna was used to receive the scattered fields off the soil sample. LO frequency was set to 2.333GHz. Rabi frequencies and LO power was identical to the previous setup. The soil sample was composed of sand (fine, particle size <0.42mm) in a 1.5x2x1.5 feet container. MW absorber foam was used to attenuate the reflections of the edges of the container. An in-situ SM probe was used to obtain ground truth (GT) data. The measurements were collected initially as a dry sample, where average SM GT was 7.95%. Tap water was added to progressively increase the volumetric SM content (VSM) to average values of 14.65%, 21.75%, and visible saturation at 34%. GT measurements were highly variable and repeated to develop error bars in Fig. 4c. Measurements at GT = 34% were normalized relative to scattering coefficient calculations from an integral equation method (I2EM)[24], which enable conversion to scattering coefficients (see Supplementary Section 3). Using a linear radar scattering coefficient to SM model[25] ($\sigma_{soil}^{dB} = \alpha M_v + \beta$, where $\alpha$ is the slope between SM and radar scattering coefficients, $\beta$ a constant dependent of soil roughness, and $M_v$ the VSM), $\alpha$ and $\beta$ is found via best fit to data, and the resulting inversion to VSM shows an error <0.5% (see Fig. 4c) – demonstrating effective SM retrieval using the atomic system over the VSM range from 7.95-34%.

**Soil moisture sensing using XM satellite signals**

SM sensing with satellite signals was achieved initially using a similar (container) concept as in Fig. 4 in an outdoor environment to utilize XM satellite signals (see Fig. 5a). The Rydberg readout and reference signal were obtained using the setup in Fig. 3d, b, respectively. The VSM change was induced progressively by adding water, and SM dynamically inverted by correlating the quantum and reference signals and using the linear SM model (see Supplementary Section 3). The GT data was obtained when the sample was dry (5%) and saturated (37%). The system (LO) was tuned to a center frequency 2.333465GHz, which is the center frequency of the XM-3 A1 sub-channel, and this channel was used for remote sensing of SM. The approach to adding tap water is highlighted in Supplementary Section 3. After each added water (AW) cycle (Fig. 5b), the SM retrievals peak initially due to surface water, then rapidly reduce to steady state SM due to flow of water. The rate of water flow is related to percolation time[26] (or hydraulic conductivity), and an exponential fit of the data after-AW cycles reveal that the percolation time constant increases with each AW cycle up to about >9 minutes near saturation (Fig. 5c). (An inverse relationship between hydraulic conductivity and the percolation rate is observed, see Supplementary Section 4).

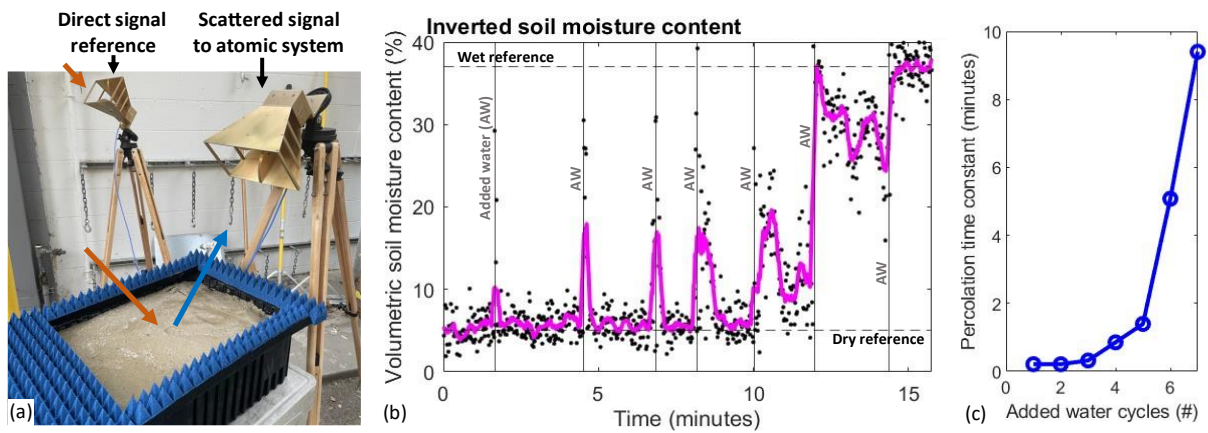

**Fig. 5: SM sensing using XM satellite signals (XM-3 A1 sub-channel). (a) The soil sample was identical to Fig. 4. (b) VSM was inverted dynamically based correlations to a reference and using a linear SM model. The solid line is a 10 point moving average (c) An exponential fit for data after each added-water (AW) cycle shows a percolation time constant that increases with each cycle to >9 minutes.**

The same configuration was used in an open outdoor terrain composed of predominantly compacted clay, with the same XM-3 A1 channel for remote sensing. The system was configured to use an additional classical readout as a means of comparing the SM retrievals from the Rydberg system (see Supplementary Section 5 for system details). In this measurement, rapid flow of water was added between 0.2-2 minutes (see Fig. 6) to the measurement site, after which the soil was visibly saturated. The measurements were collected for <15 minutes to show a slow rate of percolation due to the compacted natural soil. The Rydberg atomic SM retrievals closely matched the classical approach (see Supplementary Section 5 for more details).

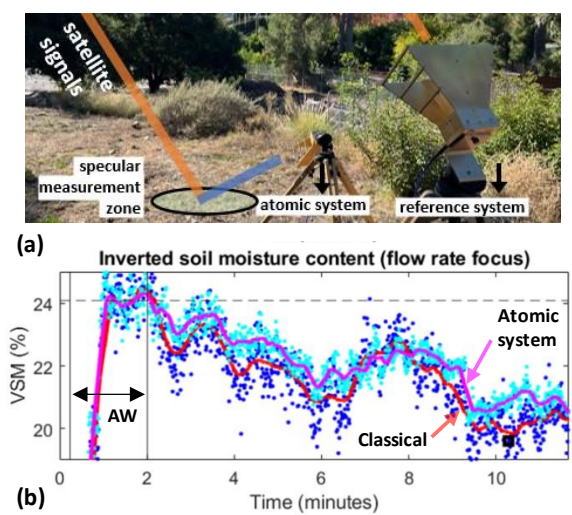

**Fig. 6: SM sensing in an outdoor natural terrain (a) and response to a rapid flow of water (duration <2minutes) with comparison to a classical SM retrieval system (b) (detailed figure in Supplementary Section 5).**

## Discussion

Spaceborne radars play a key role in remote sensing of the earth system to study natural and man-made changes in the ecosystem, water and energy cycles, weather and air quality, sea level, and surface dynamics. Different variables in the earth system (such as soil moisture, precipitation, sea surface heights, etc.) respond to different radar wavelengths or bands spanning micro-to-millimeter waves. As a result, many distinct satellite radars covering different bands are needed today to study coupled variables of the earth system. Rydberg atoms can be sensitized to enable micro-to-millimeter wave detectors that could potentially revolutionize satellite-based remote sensing by enabling a dynamically tunable radar system. In this work, we demonstrated remote sensing of soil moisture, an important earth science variable, using Rydberg atoms and satellite signals of opportunity. This work opens the possibility of dynamic broad-spectrum remote sensing via Rydberg atoms by using existing signals of opportunity. The tunability (kHz to THz) of Rydberg atomic detectors will enable broad spectrum satellite-based remote sensing for radars and radiometers when effectively coupled to broadband directive elements such as reflectors/focusing dishes[26].

## Methods

### Theory for Rydberg readout of QPSK MW

The 4-level super-heterodyning (super-het) is achieved by using a counter propagated probe and coupling laser (see Fig. 1b) and an RF field. The Hamiltonian takes the form[27]:

$$H = \frac{\hbar}{2} \begin{bmatrix} 0 & \Omega_p & 0 & 0 \\ \Omega_p & -2\Delta_p & \Omega_c & 0 \\ 0 & \Omega_c & -2(\Delta_p + \Delta_c) & \Omega_{RF} \\ 0 & 0 & \Omega_{RF} & -2(\Delta_p + \Delta_c + \Delta_{RF}) \end{bmatrix},$$

where $\Delta_p, \Delta_c$, and $\Delta_{RF}$ are the probe, coupler, and MW detuning's, and $\Omega_p, \Omega_c$, and $\Omega_{RF}$ the Rabi frequencies associated each, respectively. The setup in Fig. 1, uses an on-resonant (close to resonance) probe and coupler, so that $\Delta_{p,c} \approx 0$. The approach to sense MW is via probe laser light spectroscopy from $|1\rangle$-$|2\rangle$, which is calculated using the density matrix component ($\rho_{21}$) obtained from the master equation[28]:

$$\dot{\rho} = \frac{\partial \rho}{\partial t} = -\frac{i}{\hbar}[H, \rho] + \mathcal{L},$$

where $\mathcal{L}$ is the Lindblad operator for the atoms decay processes. When $\Delta_{RF} \approx 0$ (on-resonance MW), the probe laser light readout gives[19,23], $P_{ON}(t) \approx \langle E_T \rangle_\tau \sim E_{LO}/2 + (E_{MW}/2)\cos(\delta t + \phi_{LO})$, where $\delta$ is the difference frequency between the LO and MW RF signals. In the $\Delta_{RF} \neq 0$ off-resonant case (see Fig. 1c), the Stark shift of the optically probed state depends on the atomic polarizability[23]: $\Omega_{OFF} = \alpha \langle E^2 \rangle_\tau / 2$, where $\alpha$ is the polarizability of the state. We use Cs atoms and optically probe $66D_{5/2}$, giving an estimated polarizability of this state[27] of $\alpha = 334.08$ MHz cm²/V². In this off-resonant case:

$$P_{OFF}(t) \propto \langle E_T^2 \rangle_\tau \frac{E_{LO}^2}{2} + E_{MW} E_{LO} \cos(\delta t + \phi_{LO}),$$

where we assume $E_{LO} \gg E_{MW}$, and that $\omega_{MW} \gg \delta$. In this case, the LO field acts as a gain mechanism to the MW field and is optimized to as high a value as possible without saturating the system. In this work, we use QPSK signals (emulated or satellite signals). The signal model for QPSK is given by[21] $s(t) = E_{MW}\cos(\omega_{MW}t + \theta_n)$ within a symbol duration, where $s(t)$ is the QPSK signal, $\theta_n = (2n-1)\pi/4$ with $n = 1,2,3,4$. This implies a high-pass ($\delta > 0$) Rydberg readout with LO phase calibration (set $\phi_{LO} = 0$) given by:

$$S_{OFF}(t) \propto E_{MW} E_{LO} \cos(\delta t + \theta_n),$$

within each symbol rate. The bandwidth of $S_{OFF}(t)$ is limited by the instantaneous bandwidth (IB) of the Rydberg detector which is dependent on primarily the coupler Rabi frequency[29] and is as high as ~10MHz (empirically demonstrated) or up to ~100MHz (theoretical shot noise limited). The bandwidth or spectral efficiency of a M-PSK signal is given by $\rho = \log_2 M /2$ (bits/s)/Hz, where M is the modulation order. For QPSK signals, the modulation rate is the approximate occupied bandwidth (BW). QPSK signals with modulation rates higher than the Rydberg IB is filtered to limited to the IB.

### Theory for correlators

Cross-correlators are used in signal processing to extract the degree of similarity between two signals in reflectometry remote sensing[9,10]. This is based on a sliding dot product between the reference waveform (REF) and the sensed waveform (R, Rydberg readout):

$$(f_R \star f_{REF})(T) = \int_{-\infty}^{\infty} f_R^*(t) f_{REF}(t+T)dt,$$

where $T$ is the displacement or lag between the two waveforms. The correlation gain (gain obtained by correlating R and REF) is given by:

$$G_C = BW_{<IB} \times \tau_I,$$

where it is implied that BW≤IB, and $\tau_I$ is the integration time given simply by the duration of the reference and signal. In SoOp approaches, typical correlation gains are $G_c < 45$dB (example XM-3 A1 signal with sub-channel BW ~2MHz and duration of 1ms would give $10\log_{10} BW\tau_I$ ~33dB. For remote sensing on the ground, the delays between REF and R are negligible and $T$~0. After applying a signal correlator, the peak defined by $(f_R \star f_{REF})(0)$ gives a measured signal-to-noise (SNR) that is improved relative to the pre-correlation SNR (defined by spectral signal-to-noise estimation at 1Hz resolution bandwidth) is given by $10\log_{10} G_c$ (dB).

### Experimental set-up

Multiple experiments were conducted to advance towards satellite based remote sensing of soil moisture. The setup for the Rydberg atomic system was identical in all cases (see Fig. 1). The vapor cell was 7.5cm in length, quartz construction with 2° angled windows. The vapor pressure at 20°C was 1.1μhPa. The 4-level configuration (Fig. 1c) is realized by $6S_{1/2}$-$6P_{3/2}$-$66D_{5/2}$, and a strong (-19dBm) MW local oscillator (LO) to drive to $67P_{3/2}$. The two lowest states were $6S_{1/2}$ F=4 and $6P_{3/2}$ F = 4, driven by the probe laser. The frequency of the probe laser was locked to the hyperfine structure via a standard saturation spectroscopy technique[30], giving a probe $\lambda_p \sim 852.35672$nm. Probe linewidth is estimated to be <150kHz. The coupler laser driven to $66D_{5/2}$ was frequency locked using a cylindrical cavity (50mm diameter x 100mm length) with finesse of $F_c$ >10k. the cavity was placed inside a vacuum housing with multiple internal shields to ensure temperature control at the cavity of <1mK/day, with a thermal time constant of about 38 hours. The Pound-Drever-Hall (PDH)[31] technique was used to lock the coupler wavelength via the electronic side-band (ESB) technique[32] and to reduce the linewidth of the coupler laser to <100Hz. The coupler $\lambda_c \sim 508.98432$nm. Probe power and beam diameter was about 60µW and 1.1 mm ($1/e^2$ diameter). Coupler power and beam diameter was about 50mW and 1.3 mm. Probe and coupler laser diodes were a commercial external cavity laser diode (ECDL). The coupler at ~509nm was achieved via frequency-doubling (second harmonic generator). The resulting Rabi frequencies were $2\pi$X6.25 MHz and $2\pi$X0.68 MHz. On-resonance MW frequency was 2.385 GHz to $67P_{3/2}$. For XM detection in the 2.320-2.345 GHz band, the system was off-resonant by $\Delta f$ = 40-65 MHz. The LO signal was driven at -19dBm.

The Cs vapor cell was located inside TEM cell (length: 390 mm, width: 100 mm, height: 62 mm, septum height: 28 mm). The TEM transmission loss up to 3 GHz was <3dB. Both MW signal and LO was coupled into opposite directions of the TEM cell (see Fig. 1). A MW circulator (maximum insertion loss < 0.7dB) was used in either side to isolate the signal and LO by >18dB in each direction. MW LO signal was generated by a RF signal generator in CW (continuous wave) mode with a phase noise of about -66 dBc/Hz at 1Hz offset. The antenna used for both the atomic and reference (classical) system were identical, a double ridge guide horn antenna (aperture 13.9x24.4cm, 20.3cm length), had a gain of about ~10dB over the XM band, and a beamwidth of 30° (H-plane). A coaxial cable of length ~12m was used for all measurements except the natural terrain experiments, where lengths were ~50m. The 12m cables had a loss of <15dB (standard coaxial), while the 50m cables had a loss of <25dB (low-loss coaxial cables). The directly coupled QPSK emulated correlation experiment (Fig. 2) and the in-lab SM retrieval via emulated XM signal generation (Fig. 4) did not utilize a low-noise-amplifiers (LNA). The other experiments utilized a cascade of two LNA's with gain and noise figures of 11.5dB and 1dB, and 20dB and 1.5dB, respectively. A bandpass filter with a 2.2-2.4GHz passband was used after the LNA (when LNA is used), with a pass-band insertion loss of 1dB and an attenuation of >20dB at DC-2GHz and 2.55-8GHz. DC blocks were used prior to circulators in either side of the TEM ports to reduce/remove DC biases that may be caused by the RF signal generator (LO generation) or the LNAs (MW signal side). Additional losses of up to 3dB was measured due to RF connectors and adapters, and cables losses for MW LO was measured at 1.5dB.

For classical down-converted signals, is obtained using the same components (antenna, LNA, filter, long coax) with a RF mixer (Level 13 with <6dB conversion loss), IF (intermediate) filter and amplifier.

The soil sample used in the container experiments (Fig. 4 and 5) were composed of 100% fine sand particulates with a particle size of <0.42mm. The sample container was 1.5x2x1.5 ft. MW absorber foam was placed outside the specular region to attenuate reflections of the container and ground. The absorber attenuated reflections by >10dB at the XM band used in the measurements. Water was added evenly in the using a secondary container to reduce the flow of water via a sparse array of pinholes (container) or a fine column spray (natural terrain). More information on the measurement approach and natural terrain experiment is presented in the Supplementary Section 3 and 5. The in-situ SM probe used was a commercially available handheld RF impedance probe that operated with an CW RF frequency at 50MHz.

**Roadmap to broad-spectrum remote sensing**

To enable broad-spectrum remote sensing via SoOp, it would be necessary to eliminate classical antenna systems that are generally narrow-band and low-gain. This can most directly be achieved by coupling the vapor cell to a reflector dish, which focuses the incoming signal into the vapor cell via either its' primary focus or by using a secondary focus such as in Cassegrain or offset secondary reflector[33]. The reflector focusing dish (typically parabolic) can enable beam steering, which is convenient to dynamically alter remote sensing coverage beams and are mature for deployable applications[33]. In addition, parabolic focusing dishes have an upper frequency limited by surface roughness that can exceed extend to millimeter waves[33]. They have a gain of $G_A = e_A(\pi d/\lambda_{MW})^2$, where $e_A$ is the aperture efficiency (which is typically between 0.55-0.7, but can reach ~0.9[34]), $d$ the diameter, and $\lambda_{MW}$ the MW wavelength. As example, a deployable 1m scale focusing dish with $e_A$=0.9 and surface roughness <0.1mm, would give a gain between 20-66dB (at 2-200GHz). In addition, techniques for resonators such as the split-ring-resonators[35] can be used to further enhance fields for the low-frequencies ($\lesssim$4 GHz), however methods to integrate these with vapor cells and focusing reflectors are needed. An alternate approach is to advance arrays of Rydberg vapor cell as arrayed detectors[36], however there are considerable challenges in this approach such as coupling laser power requirements growing per node of the array.

**Figure Legends:**

**Fig. 1:** (a) High level concept for remote sensing of soil moisture based on reflected satellite signals of opportunity sensed via Rydberg atoms. (b) The probe and coupler laser light are counter-propagated in a Cesium vapor cell to excite atoms to the Rydberg state. The vapor cell is located inside a transverse electromagnetic (TEM) cell, which is used for uniform microwave (MW) field coupling to the vapor cell. Both the MW signal to be detected and the local oscillator (LO) field is coupled into the TEM cell. A beam displacer is used to develop a reference probe beam for balanced detection to reduce technical noise. Inset shows a photo of the TEM/vapor cell. (c) The coupler drives the atoms to a principal quantum number of n = 66, and MW signal and LO is off-resonant by about 40-65 MHz from the next nearest state $66D_{5/2}$-$676P_{3/2}$ transition of 2.385 GHz.

**Fig. 2:** A QPSK (quadrature phase shift key) signal with known bitstream is injected into the TEM/vapor cell to study signal-to-noise (SNR) after correlation with a reference waveform (a). The post correlation gains from the atomic readout demonstrated a gain close to the calculated $G_c$ (correlator gain) (by -3dB) (b).

**Fig. 3:** A horn antenna is connected to a low noise amplifier (LNA), then a filter, and a long coaxial cable (50m long) to drive the atomic system via the TEM cell. (a) The horn is pointed towards the XM

satelite (see Methods) to detect XM satellite signals. (b) A classical system with the same horn, LNA, filter, and coax cable is used to drive an RF mixer, IF (intermediate) filter and amplifier. Both quantum and classical readouts are digitized with an analog-to-digital convertor (ADC) and corelated to compute the envelop by sweeping local oscillator frequencies (c) Continuos readout of the XM satellite band using the correlations between (a,b). Both satellite and terrestrial repeaters are sensed. (d) The quantum systems (antenna) is directed downward towars the specular reflection to sense reflections off the soil sample (see Methods).

**Fig. 4:** An emulated XM signal (2 MHz bandwidth QPSK signal with random bitstream) is used as a transmitter to study sensitivity of Rydberg atomic detectors to soil moisture (SM). (a) The transmitter is driven by a vector signal generator. (b) A soil sample (sand, fine particle size <0.42mm) is used in a sample container of 1.5x2x1.5 ft. MW absorber foam is placed outside the specular region to attenuate reflections of the container and ground. An in-situ SM probe is used to obtain ground truth measurements. (c) Radar scattering coefficients sensed by the Rydberg atomic readout (top) and error is SM (bottom) inversion as a function of volumetric SM content, showing <0.5% error in SM inversion (see Methods for details).

**Fig. 5:** SM sensing using XM satellite signals (XM-3 A1 sub-channel). (a) The soil sample was identical to Fig. 4. (b) VSM was inverted dynamically based correlations to a reference and using a linear SM model. The solid line is a 10 point moving average (c) An exponential fit for data after each added-water (AW) cycle shows a percolation time constant that increases with each cycle to >9 minutes.

**Fig. 6:** SM sensing in an outdoor natural terrain (a) and response to a rapid flow of water (duration <2minutes) with comparison to a classical SM retrieval system (b) (detailed figure in Supplementary Section 5).

## Acknowledgements

We acknowledge discussions with P. Mao and D. Willey. The research was carried out at the Jet Propulsion Laboratory, California Institute of Technology, under a contract with the National Aeronautics and Space Administration (80NM0018D0004), through the Instrument Incubator Program's (IIP) Instrument Concept Development (Task Order 80NM0022F0020).

## Author contributions

D.A proposed the project. D.A. and J.P configured the quantum systems to include lasers and locking systems. J.P developed the optical setup. D.A., J.B., B.F., and S.P.M.N, developed the RF and SM systems and measurement concepts. D.A. designed the software scripts for data collection and processed the data for the figures. All authors supported data collection efforts. All authors contributed to discussions of the results and the manuscript.

## Additional information

The authors declare no competing interest.

**Supplementary Section 1**

**Signals of opportunity for use in soil moisture remote sensing from a spaceborne instrument**

Land and ocean remote sensing using SoOp, such as the GNSS-R, has been developed over the last 25 years, culminating in the selection of the Cyclone Global Navigation Satellite System (CYGNSS) tropical storm observation mission by NASA[37]. The use of GNSS signals at L-band (1.2 and 1.5 GHz) for surface soil moisture retrieval has been demonstrated by airborne measurements during the Soil Moisture Experiment 2002[38] and field campaigns in Europe[39-41]; its theoretical principle is based on the response of L-band microwave reflectivity to surface soil moisture[42]. Demonstration of spaceborne GNSS reflectivity dependence on surface soil moisture has been performed based on TechDemoSat-1 data[43] and extensive CYGNSS data. GNSS signals, however, have a few significant limitations, particularly for land remote sensing including very low signal power restriction of GNSS transmission in L- and S-bands, which inherently limits penetration into soil and forest canopies. A complementary technique is to use signals from communication satellites with high transmitting power. The satellites of interests are Orbcomm operating at VHF frequencies[44], the Navy's MUOS operating at VHF/UHF frequencies[45], S-band XM signals[46], and Intelsat signals at C-band frequencies[47]. These frequencies are sensitive to different parts of the vertical profile of soil moisture.

The penetration depth of different signals in the soil is dependent on the wavelength of the signal. While I/P-band penetrates the soil up to ~0.5 m (typical <30cm), L-band penetrates only the top 5 cm and is dependent on the vegetation canopy, and C-band barely penetrates the soil, but is sensitive to canopy moisture content. Therefore, these signals, if detected at the same time or over a small period, can provide vertical distribution of the soil moisture profile as has been modeled for VHF, UHF and L-band frequencies[48].

For coherent signaling due to specular responses, the spatial resolution of the Rydberg Radar will be limited to the first Fresnel zone[49]. At a reference LEO (lower earth orbit) of about 500km, this corresponds to a nominal integration time that is highly frequency dependent, varying from about 0.01-0.09s (the effective integration times are a trade-off between accuracy and spatial resolution). Currently achievable Rydberg sensitivities are $\sim 1 \mu V m^{-1} Hz^{-1/2}$ in free-space[50] for the 4-level super-heterodyning approach used here, implying a linear detector gain of up to 30x (<30dB) is needed (see Extended Data Table 1). This measure of gain is achievable with reflector focusing dishes and resonators (Method: Roadmap to broad-spectrum remote sensing).

For soil moisture, the space SoOp radar instrument will need to sample the various SoOps (Orbcomm, MUOS, GNSS, XM, IntelSat). Simultaneous sampling is not required due to the short revisit time expected. A standard architecture motivated by CYGNSS[37], with 6-12 satellites using only 4-8 parallel measurements, will provide a 10×10 km spatial resolution with a daily revisit time over the SMC region of interest of ±65°. The concept is based on a coordinated band measurement by all satellites, with a switching of bands for each revisit, or as needed. This implies that band switching time constants are relaxed. To reduce data loss, iterative switching at a time constant <10 min between satellites (upper bound is ~ 30min) is used to ensure that 90% of science data is available at every location within ±65°.

**Supplementary Section 2**

**Signal correlation and linearity**

After applying a signal correlator, the peak defined by $(f_R \star f_{REF})(\sim 0)$ gives a measured signal-to-noise (SNR) that is improved relative to the pre-correlation SNR by $10 \log_{10} G_c$ (dB). This assumes that the noise-free signal and reference would ideally have a correlation of unity. Deviations between signal and reference waveforms will reduce the post correlation SNR. Measurements in Fig. 2 show a reduction on the order of ~3dB. Extended Data Fig 1. shows the atomic down-converted spectrum for the experiment highlighted in Fig. 2. Three 10kHz QPSK modulated data at a $\Delta f$=50MHz (see Fig. 1c) spectral data-sets are shown with pre-correlation SNR's of 30, 10, and estimated -20dB.

| Value | Unit | Orbcomm | MUOS VHF | MUOS UHF | GNSS | XM | IntelSat | DTV (Ku) | DTV (K) | AEHF (K) |
|---|---|---|---|---|---|---|---|---|---|---|
| SoOp Transmitter Altitude | k-km | 0.825 | 36 | 36 | 21 | 36 | 36 | 36 | 36 | 36 |
| Center Frequency | MHz | 136.5 | 260 | 370 | 1575.4 | 2342.2 | 3950 | 12450 | 18500 | 20700 |
| Total/Min. Sub-Channel BW | MHz | 1/<1 | 20/5 | 20/5 | 1/1 | 1.8/1.8 | 500/<1 | 500/<10 | 500/<10 | 1000/<10 |
| EIRP or Transmitting Power | dBW | 12 | 27 | 43 | 26 | 68.5 | 36 | 50 | 53 | 50 |
| Total Path loss | dB/m² | 136 | 162.6 | 162.6 | 158 | 162.6 | 162.6 | 162.6 | 162.6 | 162.6 |
| Field strength at detector | µVm⁻¹ | 0.69 | 0.18 | 1.14 | 0.27 | 21.54 | 0.51 | 2.56 | 3.62 | 2.56 |
| Nominal integration time | s | 0.067 | 0.039 | 0.032 | 0.016 | 0.013 | 0.010 | 0.006 | 0.005 | 0.004 |
| **Free-space sensitivity** | **µVm⁻¹Hz⁻¹/²** | **0.18** | **0.04** | **0.21** | **0.03** | **2.45** | **0.05** | **0.19** | **0.25** | **0.17** |

**Extended Data Table 1: SoOp parameters and detector requirements**

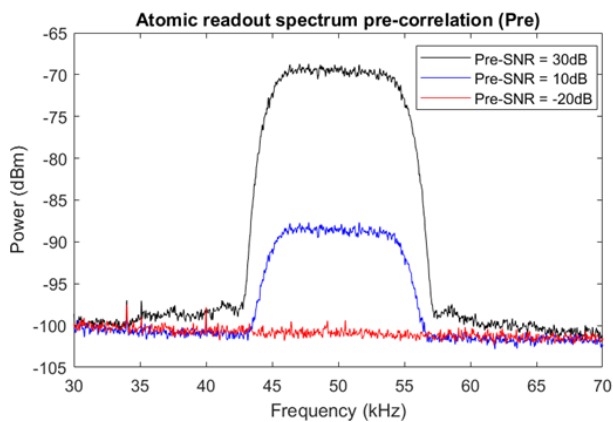

**Extended Data Fig. 1:** QPSK (quadrature phase shift keying) modulated microwave signal detected via the atomic readout. The balanced photodetector output is digitized by a base-band ADC. Result for cases above and below the pre-correlation noise level is shown (signal power is varied). QPSK signals are driven with a 10 kHz modulation rate.

The error in linearity (deviation from expected trend) of the post-correlation signal power relative to pre-correlation power is shown in Extended Data Fig. 2.

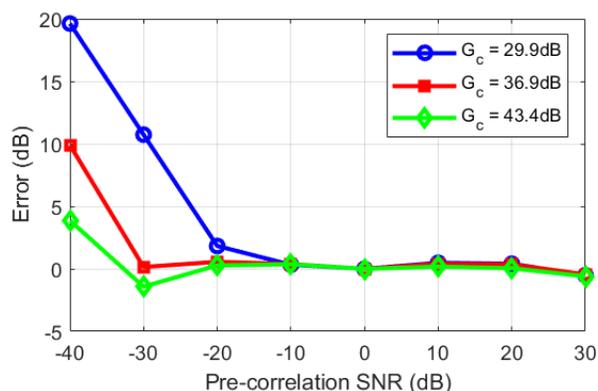

**Extended Data Fig. 2:** Linearity of post correlation peak.

**Supplementary Section 3**

**Soil moisture measurement and retrieval approach**

I2EM simulations for smooth surfaces in the specular scattering regime[51], the first order separation between roughness and moisture effects are given by[52]:

$$\sigma_{soil} = f(R)g(M_v),$$

where f is a function of roughness and g is a function of soil moisture. In log scale, this is given by[52]:

$$\sigma_{soil}^{dB} = 10\log_{10} f(R) + 10\log_{10} g(M_v)$$

Measurements show that for the same terrain, surface roughness effect can be calibrated out if no significant variation is roughness exists during the experiment[53-55], which motivates a simplified relationship between scattering and soil moisture:

$$\sigma_{soil}^{dB} = \alpha M_v + \beta,$$

where $\alpha$ is the slope between SM and radar scattering coefficients, $\beta$ a constant dependent of soil roughness, and $M_v$ (the VSM), and $\alpha$ and $\beta$ is estimated via best fit to data. Once $\alpha$ and $\beta$ are known, a direct inversion is employed between $\sigma_{soil}^{dB}$ and $M_v$.

A simple approach is taken to add water progressively for the experiments with sand samples in a container: Controlled amount of water added each time to a bucket with pre-drilled holes to control rate; Bucket location and height varied; Data collected about 15 seconds after water adding process completed (Fig. 4 only); Correlation gain of about 33.9 dB used for correlator in processing; Readout (and archive) repeated 5 times to obtain error bars / measurement uncertainty (for data collected in Fig. 4); Total data collection duration for each run is <1 minute (Fig. 4); Ground truth for SM is based on in-situ handheld SM probe (a few readings around test site was collected); This process is repeated a few times until saturation is visible. Extended Data Fig. 3 depicts an AW cycle.

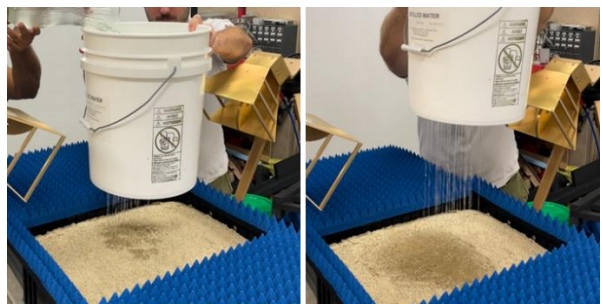

**Extended Data Fig. 3:** Adding water for container experiments.

**Supplementary Section 4**

**Percolation rate and hydraulic conductivity**

The percolation time constant was estimated (see Fig. 5c) after each AW cycle by fitting the data to an exponential function and determining the decay time constant. A 10-point moving average was used, and the fit for each AW cycle is shown in Extended Data Fig. 4.

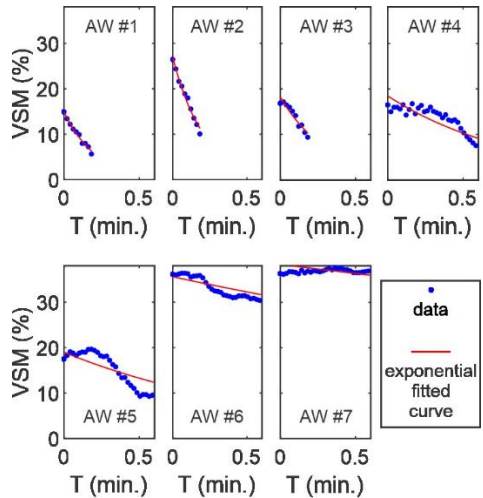

Extended Data Fig. 4: Exponential fit to determine percolation time constant after each AW cycle.

The percolation time constant (see Fig. 5c) is shown to increase gradually to >9 minutes.

Soils store and transport water to greater depths, and the ability to store and transport is determined by the hydraulic conductivity. The hydraulic conductivity is found by solving the Richards soil dynamics equation[56], and is given by[57] (neglecting sharp air-inlets):

$$\frac{k_f}{k_s} = S_e^\gamma \cdot \left[1 - \left(1 - S_e^{\frac{1}{m}}\right)^m\right]^2, \quad S_e = \frac{\theta - \theta_r}{\theta_s - \theta_r},$$

where $k_f$ is the hydraulic conductivity at current water content, $k_s$ is the hydraulic conductivity at saturation, $\gamma$ is a representation of the tortuosity and is typically set to $\gamma = 0.5$, $m = \lambda/(\lambda + 1)$, $\lambda$ the index of pore size, $\theta$ the current soil moisture content, $\theta_r$ the residual water content, and $\theta_s$ the saturation water content. We estimate $\lambda$=4124, use a $k_s$=8.28 mm/minute, and a $\theta_r$~1% to invert for $k_s - k_f$ to give the top layer steady-state value, which is shown in Extended Data Fig. 5. After each AW cycle, the percolation time constant and the hydraulic conductivity in steady state follows an inverse relation, as we would expect from normal storage and transport of water (Extended Data Fig. 6).

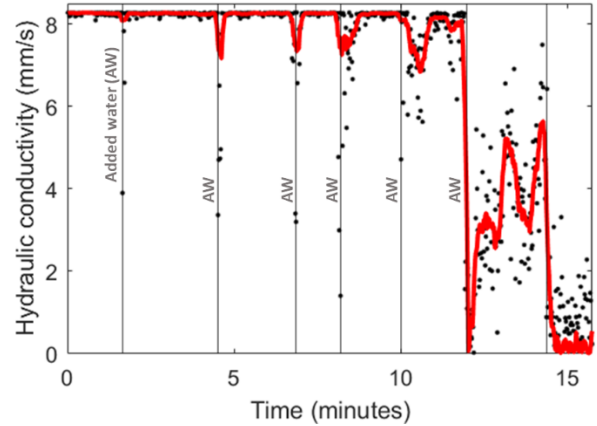

Extended Data Fig. 5: Inverted hydraulic conductivity. Solid line is a 10-point moving average.

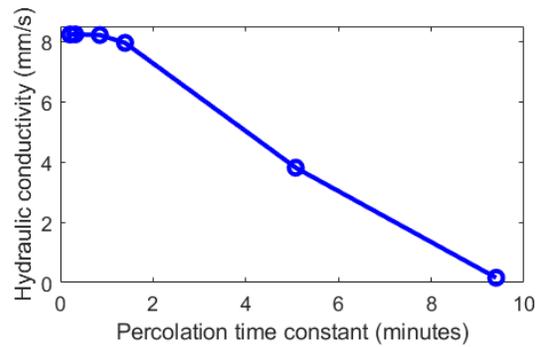

Extended Data Fig. 6: Relationship between hydraulic conductivity and percolation time constants.

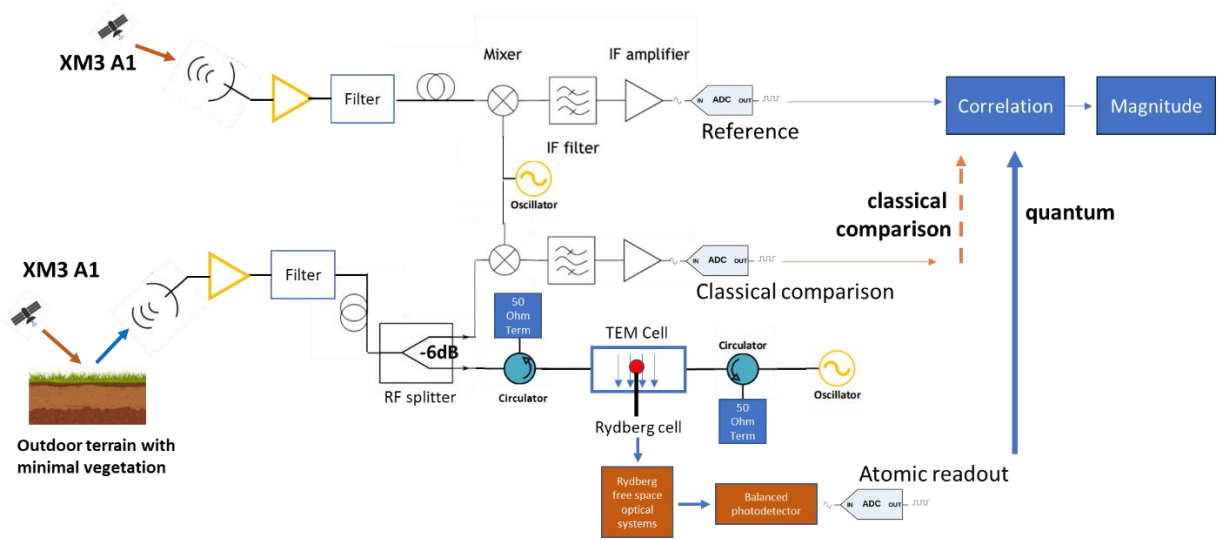

Extended Data Fig. 7: Natural terrain system architecture with an RF splitter for scattered fields to obtain a classical reflectometry readout to compare with the atomic system interpreted soil moisture data.

### Supplementary Section 5
### Experiment on natural terrain

To study the difference between a classical radar reflectometry and the atomic reflectometry system, we use an RF splitter to split (at a loss of 6dB to each) the backscatter to both the Rydberg atomic system and a classical readout system (mixer, low-pass filter baseband LNA) (see Extended Data Fig. 7). The readout from both were correlated using the approach described in the main text and compared in Fig. 6. Extended Data Fig. 8 shows a zoomed in version (bottom) of the same result to show the similarities between the classical radar reflectometry and the Rydberg atomic reflectometry results for inverted SM.

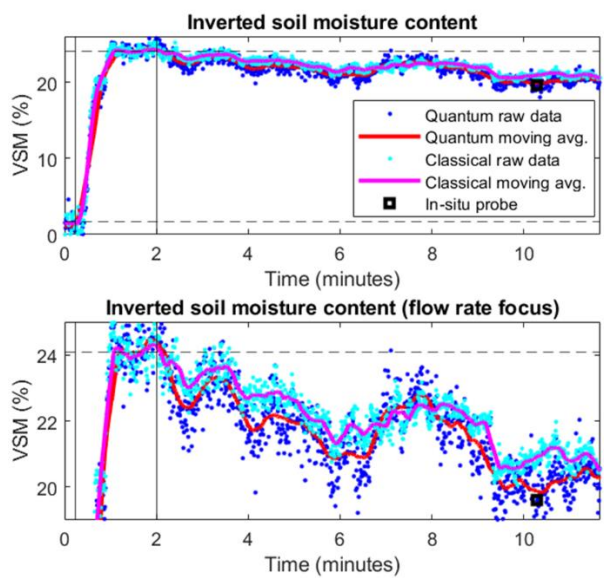

**Extended Data Fig. 8:** Flow rate focus (zoomed in) depicting the slow percolation rates in the natural terrain inverted data. The classical and quantum reflectometry results show similar trends, with minor biases due to calibrations.

### Supplementary Section References